\documentclass[10pt,12pt,onecolumn,draftclsnofoot]{IEEEtran}
\usepackage[T1]{fontenc}
\usepackage[latin9]{inputenc}
\usepackage[active]{srcltx}
\usepackage{amsmath}
\usepackage{amssymb}
\usepackage{graphicx}
\usepackage{setspace}
\usepackage[unicode=true,
 bookmarks=true,bookmarksnumbered=true,bookmarksopen=true,bookmarksopenlevel=1,
 breaklinks=false,pdfborder={0 0 0},pdfborderstyle={},backref=false,colorlinks=false]
 {hyperref}
\hypersetup{pdftitle={Your Title},
 pdfauthor={Your Name},
 pdfpagelayout=OneColumn, pdfnewwindow=true, pdfstartview=XYZ, plainpages=false}

\makeatletter
\usepackage[caption=false,font=footnotesize]{subfig}

\@ifundefined{showcaptionsetup}{}{%
 \PassOptionsToPackage{caption=false}{subfig}}
\usepackage{subfig}
\makeatother

\begin{document}
\title{Energy Efficiency Maximization for Full-Duplex UAV Secrecy Communication}
\author{Bin Duo, Qingqing Wu, Xiaojun Yuan, and Rui Zhang\thanks{B. Duo and X. Yuan are with the National Key Laboratory of Science
and Technology on Communications, University of Electronic Science
and Technology of China, Chengdu 611731, China (email: duo\_bin@163.com;
xjyuan@uestc.edu.cn). B. Duo is also with the College of Information
Science \& Technology, Chengdu University of Technology, Chengdu 610059,
China. Q. Wu and R. Zhang are with the Department of Electrical and
Computer Engineering, National University of Singapore, Singapore
117583 (e-mail: elewuqq, elezhang@nus.edu.sg).}}
\maketitle
\begin{abstract}
This letter proposes a new full-duplex (FD) secrecy communication
scheme for the unmanned aerial vehicle (UAV) and investigates its
optimal design to achieve the maximum energy efficiency (EE) of the
UAV. Specifically, the UAV receives the confidential information from
a ground source and meanwhile sends jamming signals to interfere with
a potential ground eavesdropper. As the UAV has limited on-board energy
in practice, we aim to maximize the EE for its secrecy communication,
by jointly optimizing the UAV trajectory and the source/UAV transmit/jamming
powers over a finite flight period with given initial and final locations.
Although the problem is difficult to solve, we propose an efficient
iterative algorithm to obtain its suboptimal solution.\emph{ }Simulation
results show that the proposed joint design can significantly improve
the EE of UAV secrecy communication, as compared to various benchmark
schemes.
\end{abstract}


\IEEEpeerreviewmaketitle{}

\section{Introduction}

Unmanned aerial vehicles (UAVs) have been widely used in wireless
communications, thanks to their line-of-sight (LoS) air-to-ground
links and high controllable mobility \cite{Zeng:2019tl}. Despite
their numerous applications, UAV communications face new challenges.
Among others, due to the LoS links, the legitimate UAV communications
are more prone to the interception by suspicious eavesdroppers on
the ground \cite{Wu:2019sa}. Fortunately, the high mobility of UAVs
provides a new opportunity to enhance the secrecy rate of UAV communications
by leveraging proper trajectory design. The authors in \cite{Zhang:2019jq}
and \cite{Li:2018ix} study secure UAV communications, where the average
secrecy rate is significantly improved by jointly optimizing the UAV
trajectory and power allocation over a given mission duration. Dual-UAV
systems are proposed in \cite{Cai:2018je} and \cite{Zhong:2018bf},
where one UAV communicates with legitimate ground users while the
other UAV safeguards their transmission by cooperatively sending jamming
signals to ground eavesdroppers. In addition, the propulsion energy
required for the UAVs to keep airborne and enable high mobility is
practically limited due to their finite on-board energy and thus needs
to be taken into account for the trajectory design \cite{Zeng:2018ux}.
Therefore, the energy efficiency (EE) of fixed-wing UAVs has been
maximized in \cite{Xiao:2018va}, where the UAV serves as a mobile
relay to assist in the secure communication between two ground nodes.
However, different from rotary-wing UAVs, fixed-wing UAVs require
a certain minimum speed to keep airborne and thus cannot hover at
a fixed location to sustain the maximum secrecy rate \cite{Zeng:2019tl}.
Note that the above studies all consider half-duplex UAV communications
under the secrecy setup. As such, it remains unaddressed whether full-duplex
(FD) UAV communications can further improve the secrecy rate.

Motivated by the above, this letter investigates the EE-optimal joint
secrecy communication and trajectory design for rotary-wing UAVs by
exploiting the FD communication at the UAV. Specifically, we consider
a scenario where a flying UAV intends to receive confidential information
from a ground source and in the meanwhile avoid information leakage
to a suspicious eavesdropper on the ground by sending jamming signals
to interfere with it. To balance between the secrecy rate and the
energy consumption of the UAV, we aim to maximize the EE for the UAV
secrecy communication by jointly optimizing the UAV trajectory and
the source/UAV transmit/jamming power allocations along its trajectory.
In the proposed design, the UAV is subjected to its mobility as well
as both the average and peak transmit power constraints. To resolve
the non-convexity of the formulated problem, we propose an efficient
iterative algorithm to obtain a high-quality suboptimal solution for
it based on the block coordinate descent (BCD) and successive convex
approximation (SCA) techniques. Simulation results validate that the
proposed joint design significantly improves the EE of the UAV, as
compared to other benchmark schemes without FD transmission, power
control, or trajectory optimization.

\begin{figure}[t]
\begin{centering}
\includegraphics[width=7cm,height=4.5cm]{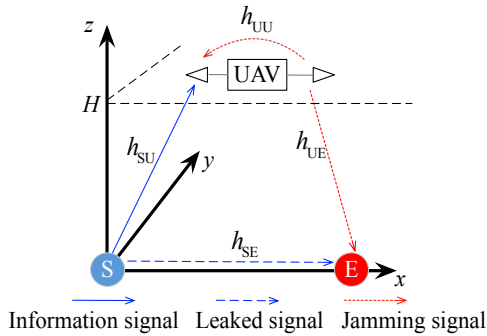}
\par\end{centering}
\centering{}\caption{A full-duplex UAV secrecy communication system.}
\end{figure}

\section{System Model and Problem Formulation}

As shown in Fig. 1, we consider a UAV-enabled wireless communication
system, where a ground source (S) transmits confidential information
to a rotary-wing UAV, while a ground eavesdropper (E) tries to overhear
it. To secure the ground-to-air communication, we assume that the
UAV operates in FD mode under which it can send jamming signals to
interfere with E while receiving the secrecy information from S. Without
loss of generality, we consider a three-dimensional (3D) Cartesian
coordinate system, where S and E are located on the ground with horizontal
coordinates $[0,0]^{T}$ and $\mathbf{w}_{\textrm{E}}=[x_{\textrm{E}},y_{\textrm{E}}]^{T}$,
respectively, and their locations are assumed to be fixed and known
to the UAV for the period of interest. Let $T>0$ and $[x(t),y(t)]^{T}$
respectively denote a given finite flight period of the UAV and its
horizontal coordinate at each time instant $t$, $0\leq t\leq T$,
where we assume that the UAV flies at a fixed altitude denoted by
$H$. Similar to \cite{Zhang:2019jq}, $T$ is divided into $N$ time
slots with equal length, i.e., $T=\delta_{t}N$, where $\delta_{t}$
denotes the duration of each time slot and is practically set sufficiently
small. As such, the UAV\textquoteright s horizontal trajectory over
$T$ can be represented approximately by a sequence of locations denoted
by $\mathbf{q}=\{\mathbf{q}[n]\triangleq\left[x[n],y[n]\right]^{T}\}$,
with $x[n]=x(n\delta_{t}),y[n]=y(n\delta_{t}),n=0,\ldots,N$. We assume
that the UAV 's initial and final locations are given by $\mathbf{q}_{0}=[x_{0},y_{0}]^{T}$
and $\mathbf{q}_{\textrm{F}}=[x_{\textrm{F}},y_{\textrm{F}}]^{T}$,
respectively. Let the maximum speed of the UAV be $V_{\max}$ in meter/second
(m/s) and thus $\varOmega=V_{\max}\delta_{t}$ is the maximum horizontal
distance that the UAV can fly within each time slot. Then, the UAV
trajectory needs to satisfy the following constraints: 
\begin{equation}
||\mathbf{q}[n+1]-\mathbf{q}[n]||^{2}\leq\varOmega^{2},n=0,\cdots,N-1,\label{eq:mobility}
\end{equation}
\begin{equation}
\mathbf{q}[0]=\mathbf{q}_{0},\mathbf{q}[N]=\mathbf{q}_{\textrm{F}}.\label{eq:initial and final}
\end{equation}

We assume that the transmission from S to the UAV and that from the
UAV to E are both dominated by LoS channels {[}3{]}-{[}8{]}. Thus,
the corresponding channel power gains in time slot $n$ follow the
free-space path loss model, given by $h_{\textrm{SU}}[n]=\rho_{0}/(H^{2}+||\mathbf{q}[n]||^{2})$
and $h_{\textrm{UE}}[n]=\rho_{0}/(H^{2}+||\mathbf{q}[n]-\mathbf{w}_{\textrm{E}}||^{2})$,
respectively, where $\rho_{0}$ denotes the channel power gain at
the reference distance $d_{0}=1$ m. The terrestrial channel between
S and E is assumed to follow Rayleigh fading with the channel power
gain denoted by $h_{\textrm{SE}}=\rho_{0}||\mathbf{w}_{\textrm{E}}||{}^{-\kappa}\zeta$,
where $\kappa\geq2$ is the path-loss exponent and $\zeta$ is an
exponentially distributed random variable with unit mean accounting
for small-scale Rayleigh fading. Since the residual self-interference
(RSI) is difficult to be completely removed in practice for FD radios,
we take into account its impact on the UAV secrecy communication performance.
Let $h_{\textrm{UU}}$ denote the channel gain that characterizes
the RSI due to imperfect loop interference cancellation from the UAV's
transmitting antenna to its receiving antenna. The RSI channel $h_{\textrm{UU}}$
is commonly modeled as Rayleigh fading, i.e., $h_{\textrm{UU}}$ is
independently drawn from $\mathcal{CN}(0,\sigma_{\textrm{RSI}}^{2})$,
where $\sigma_{\textrm{RSI}}^{2}$ is regarded as the average loop
interference level (LIL)\emph{ }with $\mathbb{E}[|h_{\textrm{UU}}|^{2}]=\sigma_{\textrm{RSI}}^{2}$
\cite{Ngo:2014jq}\emph{.}

Let $p_{\textrm{S}}[n]$ and $p_{\textrm{U}}[n]$ denote the source
transmit power and the UAV jamming power in time slot $n$, respectively.
In practice, they should satisfy both the average and peak power constraints
given as follows 
\begin{equation}
\frac{1}{N}\sum_{n=1}^{N}p_{\textrm{S}}[n]\leq\bar{P}_{\textrm{S}},0\leq p_{\textrm{S}}[n]\leq P_{\textrm{S}}^{\max},\label{eq:Ps}
\end{equation}
\begin{equation}
\frac{1}{N}\sum_{n=1}^{N}p_{\textrm{U}}[n]\leq\bar{P}_{\textrm{U}},0\leq p_{\textrm{U}}[n]\leq P_{\textrm{U}}^{\max},\label{eq:Pu}
\end{equation}
where $\bar{P}_{\textrm{S}}\leq P_{\textrm{S}}^{\max}$ and $\bar{P}_{\textrm{U}}\leq P_{\textrm{U}}^{\max}$.
Then, the achievable rates in bits/second/Hertz (bps/Hz) of the UAV
and the eavesdropper in time slot $n$ are respectively given by 
\begin{align}
R_{\textrm{U}}[n] & =\mathbb{E}_{h_{\textrm{UU}}}\left[\log_{2}\left(1+\frac{p_{\textrm{S}}[n]h_{\textrm{SU}}[n]}{p_{\textrm{U}}[n]|h_{\textrm{UU}}|^{2}+\sigma^{2}}\right)\right]\nonumber \\
 & \overset{(a)}{\geq}\log_{2}\left(1+\frac{p_{\textrm{S}}[n]h_{\textrm{SU}}[n]}{p_{\textrm{U}}[n]\sigma_{\textrm{RSI}}^{2}+\sigma^{2}}\right)\triangleq\check{R}{}_{\textrm{U}}[n],
\end{align}
\begin{align}
R_{\textrm{E}}[n] & =\mathbb{E}_{\zeta}\left[\log_{2}\left(1+\frac{p_{\textrm{S}}[n]h_{\textrm{SE}}}{p_{\textrm{U}}[n]h_{\textrm{UE}}[n]+\sigma^{2}}\right)\right]\nonumber \\
 & \overset{(b)}{\leq}\log_{2}\left(1+\frac{p_{\textrm{S}}[n]\rho_{0}||\mathbf{w}_{\textrm{E}}||{}^{-\kappa}}{p_{\textrm{U}}[n]h_{\textrm{UE}}[n]+\sigma^{2}}\right)\triangleq\hat{R}{}_{\textrm{E}}[n],\label{eq:Re_ub}
\end{align}
where $\mathbb{E}_{h_{\textrm{UU}}}[\cdot]$ and $\mathbb{E}_{\zeta}[\cdot]$
are the expectation operators with respect to $h_{\textrm{UU}}$ and
$\zeta$, respectively, and $\sigma^{2}$ is the additive white Gaussian
noise power at the corresponding receiver. Note that due to the convexity
of $R_{\textrm{U}}[n]$ and concavity of $R_{\textrm{E}}[n]$ with
respect to the corresponding random variables, $(a)$ in (5) and $(b)$
in (6) hold based on Jensen's inequality. Hence, the achievable secrecy
rate for each time slot $n$ is lower-bounded by 
\begin{equation}
R_{\textrm{sec}}[n]=\left[\check{R}{}_{\textrm{U}}[n]-\hat{R}{}_{\textrm{E}}[n]\right]^{+},\label{eq:Rsec}
\end{equation}
where $[x]^{+}=\max(x,0)$. Note that the operation $[\cdot]^{+}$
can be dropped since the practical value of (\ref{eq:Rsec}) is at
least zero by setting $p_{\textrm{S}}[n]=0$ for any $n$.

In practice, the communication-related energy is much smaller than
the propulsion energy of UAVs, and thus is ignored in this letter.
Based on \cite{Zeng:2018ux}, the propulsion energy consumption $E_{\textrm{p}}[n]$
in Joule (J) for rotary-wing UAVs with speed $v[n]$ in time slot
$n$ can be modeled as 
\begin{equation}
E_{\textrm{p}}[n]=\delta_{t}(P_{0}\phi[n]+P_{i}\varphi^{1/2}[n]+\frac{1}{2}d_{0}\rho sAv^{3}[n]),\label{eq:E(V)}
\end{equation}
where the UAV (horizontal) flying speed is given by\emph{ $v[n]=||\mathbf{q}[n+1]-\mathbf{q}[n]||/\delta_{t}$,
$\phi[n]=1+3v^{2}[n]/U_{tip}^{2}$}, $\varphi[n]=(1+v^{4}[n]/(4v_{0}^{4}))^{1/2}-v^{2}[n]/(2v_{0}^{2})$,
$P_{0}$, $P_{i}$ and \emph{$v_{0}$ }are constants, which represent
the blade profile power, induced power, and the mean rotor induced
speed when the UAV is hovering, respectively, $U_{tip}$ is the tip
speed of the UAV's rotor blade, $s$ and $d_{0}$ denote the rotor
solidity and fuselage drag ratio, respectively, and $A$ and $\rho$
are the rotor disc area and air density, respectively. Note that (\ref{eq:E(V)})
is practically valid for the straight and level flight of rotary-wing
UAVs, which is satisfied in each time slot $n$ due to the approximated
piecewise-linear UAV trajectory over time slots.

We aim to maximize the EE for the UAV secrecy communication in bits/J
over $N$ time slots by jointly optimizing the source transmit power
$\mathbf{p}_{\textrm{S}}\triangleq\left\{ p_{\textrm{S}}[n]\right\} _{n=1}^{N}$,
the UAV jamming power $\mathbf{p}_{\textrm{U}}\triangleq\left\{ p_{\textrm{U}}[n]\right\} _{n=1}^{N}$
and the UAV trajectory $\mathbf{q}$. This optimization problem can
be formulated as
\begin{align}
 & \max_{\mathbf{p}_{\textrm{S}},\mathbf{p}_{\textrm{U}},\mathbf{q}}\frac{B\sum_{n=1}^{N}R_{\textrm{sec}}[n]}{\sum_{n=1}^{N}E_{\textrm{p}}[n]}\label{eq:P1}\\
 & \quad\textrm{s.t.}\;(\ref{eq:mobility})-(\ref{eq:Pu}),\nonumber 
\end{align}
where $B$ denotes the system bandwidth. Problem (\ref{eq:P1}) is
difficult to be optimally solved in general since the objective function
is not jointly concave with respect to the optimization variables.

\section{Proposed Algorithm}

In this section, we propose an efficient iterative algorithm to obtain
a high-quality suboptimal solution to problem (\ref{eq:P1}) by applying
BCD and SCA methods. Specifically, problem (\ref{eq:P1}) is tackled
by iteratively solving three subproblems to optimize each of the source
transmit power $\mathbf{p}_{\textrm{S}}$, the UAV jamming power $\mathbf{p}_{\textrm{U}}$,
and the UAV trajectory $\mathbf{q}$ with the other two being fixed,
until the algorithm converges.

\subsection{Source Power Optimization}

For any given jamming power $\mathbf{p}_{\textrm{U}}$ and UAV trajectory
$\mathbf{q}$, problem (\ref{eq:P1}) is reduced to 
\begin{align}
 & \max_{\mathbf{p}_{\textrm{S}}}\sum_{n=1}^{N}\left[\log_{2}\left(1+a_{n}p_{\textrm{S}}[n]\right)-\log_{2}\left(1+b_{n}p_{\textrm{S}}[n]\right)\right]\label{eq:P3}\\
 & \quad\textrm{s.t.}\;(\ref{eq:Ps}),\nonumber 
\end{align}
where $a_{n}=\gamma_{0}/((H^{2}+||\mathbf{q}[n]||^{2})(p_{\textrm{U}}[n]\beta_{0}+1))$,
$b_{n}=\gamma_{0}||\mathbf{w}_{\textrm{E}}||{}^{-\kappa}/(\frac{\gamma_{0}p_{\textrm{U}}[n]}{H^{2}+||\mathbf{q}[n]-\mathbf{w}_{\textrm{E}}||^{2}}+1)$,
$\gamma_{0}=\rho_{0}/\sigma^{2}$ is the reference signal-to-noise
ratio (SNR), and $\beta_{0}=\sigma_{\textrm{RSI}}^{2}/\sigma^{2}$
is defined as the LIL-to-noise ratio. According to \cite{Zhang:2019jq},
the optimal solution is given by $p_{\textrm{S}}^{*}[n]=\min([\eta_{n}]^{+},P_{\textrm{S}}^{\max})$
if $a_{n}>b_{n}$; otherwise $p_{\textrm{S}}^{*}[n]=0$, where $\eta_{n}=[(1/(2b_{n})-1/(2a_{n}))^{2}+(1/b_{n}-1/a_{n})/(\mu\ln2)]^{\frac{1}{2}}-1/(2a_{n})-1/(2b_{n})$.
Note that $\mu\geq0$ is a constant that ensures $\sum_{n=1}^{N}p_{\textrm{S}}^{*}[n]\leq N\bar{P}_{\textrm{S}}$,
which can be obtained efficiently via the bisection method.

\subsection{Jamming Power Optimization}

To solve this subproblem for any given $\mathbf{p}_{\textrm{S}}$
and $\mathbf{q}$, we notice that each term in $R_{\textrm{sec}}[n]$
can be expressed by a difference of two concave functions with respect
to $\mathbf{p}_{\textrm{U}}$, i.e., 
\begin{align}
R_{\textrm{sec}}[n] & =\log_{2}\left(\beta_{0}p_{\textrm{U}}[n]+1+c_{n}\right)-\log_{2}\left(\beta_{0}p_{\textrm{U}}[n]+1\right)\nonumber \\
 & \quad-\log_{2}\left(e_{n}p_{\textrm{U}}[n]+1+d_{n}\right)+\log_{2}\left(e_{n}p_{\textrm{U}}[n]+1\right),\label{eq:11}
\end{align}
where $c_{n}=p_{\textrm{S}}[n]\gamma_{0}/(H^{2}+||\mathbf{q}[n]||^{2})$,
$d_{n}=p_{\textrm{S}}[n]\gamma_{0}||\mathbf{w}_{\textrm{E}}||{}^{-\kappa}$
and $e_{n}=\gamma_{0}/(H^{2}+||\mathbf{q}[n]-\mathbf{w}_{\textrm{E}}||^{2})$.
Despite the non-convexity of (\ref{eq:11}), we can employ the SCA
method to approximately solve it. Denote by $\mathbf{p}_{\textrm{U}}^{k}=\left\{ p_{\textrm{U}}^{k}[n]\right\} _{n=1}^{N}$
the jamming power of the UAV in the $k$-th iteration. Due to the
concavity of $\log_{2}\left(\beta_{0}p_{\textrm{U}}[n]+1\right)$
and $\log_{2}\left(e_{n}p_{\textrm{U}}[n]+1+d_{n}\right)$ in (\ref{eq:11}),
we can obtain their respective globally upper bounds by applying the
first-order Taylor expansion at $p_{\textrm{U}}^{k}[n]$, i.e., 
\begin{equation}
\log_{2}\left(\beta_{0}p_{\textrm{U}}[n]+1\right)\leq\log_{2}\left(\beta_{0}p_{\textrm{U}}^{k}[n]+1\right)+A^{k}[n],\label{eq:p2ub1}
\end{equation}
\begin{align}
\log_{2}\left(e_{n}p_{\textrm{U}}[n]+1+d_{n}\right) & \leq\log_{2}\left(e_{n}p_{\textrm{U}}^{k}[n]+1+d_{n}\right)+B^{k}[n],\label{eq:p2ub2}
\end{align}
where $A^{k}[n]=\beta_{0}(p_{\textrm{U}}[n]-p_{\textrm{U}}^{k}[n])/(\ln2(\beta_{0}p_{\textrm{U}}^{k}[n]+1))$
and $B^{k}[n]=e_{n}(p_{\textrm{U}}[n]-p_{\textrm{U}}^{k}[n])/(\ln2(e_{n}p_{\textrm{U}}^{k}[n]+1+d_{n}))$.
Based on (\ref{eq:p2ub1}) and (\ref{eq:p2ub2}), problem (\ref{eq:P1})
can be approximately reformulated as the following problem, 
\begin{align}
 & \max_{\mathbf{p}_{\textrm{U}}}\sum_{n=1}^{N}\Big[\log_{2}(\beta_{0}p_{\textrm{U}}[n]+1+c_{n})+\log_{2}(e_{n}p_{\textrm{U}}[n]+1)\nonumber \\
 & \qquad\qquad-A^{k}[n]-B^{k}[n]\Big]\label{eq:P5}\\
 & \quad\textrm{s.t.}\;(\ref{eq:Pu}),\nonumber 
\end{align}
Note that subproblem (\ref{eq:P5}) is convex and thus can be solved
efficiently by the CVX solver\emph{.} Since the upper bounds in (\ref{eq:p2ub1})
and (\ref{eq:p2ub2}) suggest that any feasible solution $\mathbf{p}_{\textrm{U}}^{k}$
to (\ref{eq:P1}) is also feasible for (\ref{eq:P5}), the optimal
value obtained by solving (\ref{eq:P5}) serves as a lower bound for
that of problem (\ref{eq:P1}).

\subsection{UAV Trajectory Optimization}

Even with given $\mathbf{p}_{\textrm{S}}$ and $\mathbf{p}_{\textrm{U}}$,
problem (9) is still difficult to be solved optimally, due to the
non-convexity of its objective function with respect to $\mathbf{q}$.
To tackle the non-convexity of $R_{\textrm{sec}}[n]$ in (\ref{eq:P1}),
we first introduce slack variables $\mathbf{g}=\left\{ g[n]\right\} _{n=1}^{N}$
and $\mathbf{m}=\left\{ m[n]\right\} _{n=1}^{N}$, where $g[n]\geq H^{2}+||\mathbf{q}[n]||^{2}$
and $m[n]\geq H^{2}+||\mathbf{q}[n]-\mathbf{w}_{\textrm{E}}||^{2}$.
Thus, $R_{\textrm{sec}}[n]$ can be written as 
\begin{equation}
R_{\textrm{sec}}[n]=\sum_{n=1}^{N}\left[\log_{2}\left(1+\frac{f_{n}}{g[n]}\right)-\log_{2}\left(1+\frac{d_{n}m[n]}{\gamma_{0}p_{\textrm{U}}[n]+m[n]}\right)\right],
\end{equation}
where $f_{n}=\gamma_{0}p_{\textrm{S}}[n]/(\beta_{0}p_{\textrm{U}}[n]+1)$.
Note that the constraints for $\mathbf{g}$ and $\mathbf{m}$ must
hold with equalities to obtain the optimal solution to problem (\ref{eq:P1}),
since otherwise $g[n]$ and $m[n]$ can be increased to decrease the
objective value. Similarly, by using the first-order Taylor expansion,
the first and second terms in (15) can be replaced by their respective
convex lower and concave upper bounds, at given local points denoted
by $\mathbf{g}^{k}=\left\{ g^{k}[n]\right\} _{n=1}^{N}$ and $\mathbf{m}^{k}=\left\{ m^{k}[n]\right\} _{n=1}^{N}$
in the $k$-th iteration. Specifically, we have 
\begin{equation}
\log_{2}\left(1+\frac{f_{n}}{g[n]}\right)\geq R_{\textrm{sec}}^{\textrm{lb}}[n],\label{eq:p6_lb}
\end{equation}
\begin{equation}
\log_{2}\left(1+\frac{d_{n}m[n]}{\gamma_{0}p_{\textrm{U}}[n]+m[n]}\right)\leq R_{\textrm{sec}}^{\textrm{ub}}[n],\label{eq:p6_ub}
\end{equation}
where $R_{\textrm{sec}}^{\textrm{lb}}[n]=\log_{2}(1+f_{n}/g^{k}[n])-f_{n}(g[n]-g^{k}[n])/(\ln2(g^{k}[n]+f_{n})g^{k}[n])$,
$R_{\textrm{sec}}^{\textrm{ub}}[n]=C^{k}[n](m[n]-m^{k}[n])+D^{k}[n]$,
$C^{k}[n]=d_{n}\gamma_{0}p_{\textrm{U}}[n]/(\ln2(\gamma_{0}p_{\textrm{U}}[n]+(d_{n}+1)m^{k}[n])(\gamma_{0}p_{\textrm{U}}[n]+m^{k}[n]))$
and $D^{k}[n]=\log_{2}(1+d_{n}m^{k}[n]/(\gamma_{0}p_{\textrm{U}}[n]+m^{k}[n]))$.

Then, to tackle the non-convexity of $E_{\textrm{p}}[n]$ in problem
(\ref{eq:P1}),\emph{ }we further introduce slack variable $\mathbf{s}=\left\{ s[n]\right\} _{n=1}^{N}$
such that $s[n]\geq[(1+v^{4}[n]/(4v_{0}^{4}))^{\frac{1}{2}}-v^{2}[n]/(2v_{0}^{2})]^{\frac{1}{2}}$,
which is equivalent to 
\begin{equation}
\frac{1}{s^{2}[n]}\leq s^{2}[n]+\frac{v^{2}[n]}{v_{0}^{2}}=s^{2}[n]+\frac{||\mathbf{q}[n+1]-\mathbf{q}[n]||^{2}}{v_{0}^{2}\delta_{t}^{2}}.\label{eq:t=00005Bn=00005D2}
\end{equation}
Note that the constraint (\ref{eq:t=00005Bn=00005D2}) should hold
with equality to obtain the optimal solution, since otherwise $s[n]$
can be increased to decrease the objective value of problem (\ref{eq:P1}).
Next, we focus on addressing\emph{ }the non-convex constraint (\ref{eq:t=00005Bn=00005D2}).
Since $s^{2}[n]$ and $||\mathbf{q}[n+1]-\mathbf{q}[n]||^{2}$ are
convex with respect to $s[n]$ and $\mathbf{q}[n]$, respectively,
we can apply the first-order Taylor expansion to the right hand side
(RHS) of (\ref{eq:t=00005Bn=00005D2}) at any given points $\mathbf{s}^{k}=\left\{ s^{k}[n]\right\} _{n=1}^{N}$
and $\mathbf{q}^{k}=\left\{ \mathbf{q}^{k}[n]\right\} _{n=1}^{N}$
in the $k$-th iteration to obtain the following lower bound, i.e.,
\begin{align}
 & s^{2}[n]+\frac{||\mathbf{q}[n+1]-\mathbf{q}[n]||^{2}}{v_{0}^{2}\delta_{t}^{2}}\geq(s^{k}[n])^{2}+2s^{k}[n](s[n]-s^{k}[n])\nonumber \\
 & \qquad\qquad-\frac{||\boldsymbol{\psi}^{k}[n]||^{2}}{v_{0}^{2}\delta_{t}^{2}}+\frac{2}{v_{0}^{2}\delta_{t}^{2}}(\boldsymbol{\psi}^{k}[n])^{T}(\mathbf{q}[n+1]-\mathbf{q}[n])\triangleq F^{k}[n],\label{eq:t=00005Bn=00005D con lb}
\end{align}
where $\boldsymbol{\psi}^{k}[n]=\mathbf{q}^{k}[n+1]-\mathbf{q}^{k}[n]$.

With (\ref{eq:p6_lb})-(\ref{eq:t=00005Bn=00005D con lb}), we obtain
the following optimization problem
\begin{align}
 & \max_{\mathbf{q},\mathbf{g},\mathbf{m},\mathbf{s}}\frac{\sum_{n=1}^{N}\left[R_{\textrm{sec}}^{\textrm{lb}}[n]-R_{\textrm{sec}}^{\textrm{ub}}[n]\right]}{\sum_{n=1}^{N}\left[P_{0}\phi[n]+P_{i}s[n]+\frac{1}{2}d_{0}\rho sAv^{3}[n]\right]}\label{eq:P7}\\
 & \quad\textrm{s.t.}\;\;\frac{1}{s^{2}[n]}\leq F^{k}[n],\\
 & \quad\qquad g[n]\geq H^{2}+||\mathbf{q}[n]||^{2},\\
 & \quad\qquad m[n]\geq H^{2}+||\mathbf{q}[n]-\mathbf{w}_{\textrm{E}}||^{2},\\
 & \quad\qquad s[n]\geq0,\\
 & \quad\qquad(\ref{eq:mobility})-(\ref{eq:initial and final}).\nonumber 
\end{align}
It is observed that problem (\ref{eq:P7}) is a quasi-convex optimization
problem since its objective function is composed of a linear numerator
and a convex denominator and all constraints are convex. As such,
it can be optimally and efficiently solved via fractional programming
techniques, e.g., the Dinkelbach\textquoteright s algorithm. Note
that the lower bound in (\ref{eq:p6_lb}) and the upper bound in (\ref{eq:p6_ub})
suggest that the feasible set of $\mathbf{g}^{k}$ and $\mathbf{m}^{k}$
for problem (\ref{eq:P7}) is always a subset of that of problem (\ref{eq:P1}).
As a result, the optimal value obtained by solving problem (\ref{eq:P7})
is a lower bound for that of problem (\ref{eq:P1}).

To sum up, we solve the three subproblems (\ref{eq:P3}), (\ref{eq:P5}),
and (\ref{eq:P7}) alternately in an iterative manner to obtain the
suboptimal solution to problem (\ref{eq:P1}) until the fractional
increase of the objective value is less than a given threshold, $\epsilon>0$.

\section{Numerical Results}

In this section, we show simulation results on comparing the proposed
joint power control and trajectory design algorithm (denoted as P\&T)
with three benchmark algorithms: 1) UAV trajectory optimization without
jamming (denoted as NJ/T); 2) UAV trajectory design without power
control (denoted as NP/T); and 3) best-effort trajectory with power
control (P/BET). Specifically\emph{,} the NJ/T algorithm jointly optimizes
the source transmit power $p_{\textrm{S}}[n]$ and the UAV trajectory
by setting $p_{\textrm{U}}[n]=0,\forall n$ in P\&T. Note that if
the UAV energy consumption is not considered, the NJ/T algorithm is
the same as that given in \cite{Zhang:2019jq}. In NP/T, the powers
of the UAV and the source are set as $p_{\textrm{U}}[n]=\bar{P}_{\textrm{U}}$
and $p_{\textrm{S}}[n]=\bar{P}_{\textrm{S}},\forall n$, respectively,
and the UAV trajectory is optimized by iteratively solving (\ref{eq:P7})
until convergence. The best-effort trajectory in P/BET is designed
as follows (not shown separately in Fig. 2 due to space limitation):
the UAV first flies along a straight line towards the location right
above the source at speed $V_{\max}$, then (if time permits) stays
stationary as long as possible, and finally flies at speed $V_{\max}$
to its final location by the end of $T$. With given best-effort trajectory
in P/BET, the powers $p_{\textrm{S}}[n]$ and $p_{\textrm{U}}[n]$
are optimized by solving problems (\ref{eq:P3}) and (\ref{eq:P5}),
respectively. The simulation parameters are set as $\mathbf{q}_{0}=[50,-800]^{T}$
m, $\mathbf{q}_{\textrm{F}}=[50,800]^{T}$ m, $\mathbf{w}_{\textrm{E}}=[200,0]^{T}$
m, $H=100$ m, $V_{\max}=40$ m/s, $\delta_{t}=0.5$ s, $\rho_{0}=-60$
dB, $\bar{P}_{\textrm{S}}=20$ dBm, $P_{\textrm{S}}^{\max}=26$ dBm,
$\bar{P}_{\textrm{U}}=10$ dBm, $P_{\textrm{U}}^{\max}=16$ dBm, $B=1$
MHz, $\sigma^{2}=-110$ dBm, and $\epsilon=10^{-4}$. The values of
all required parameters in (\ref{eq:E(V)}) are set according to the
example given in \cite{Zeng:2018ux}.

\begin{figure*}[t]
\begin{centering}
\subfloat[UAV's Trajectories by T\&P algorithm.]{\centering{}\includegraphics[width=8cm,height=6cm]{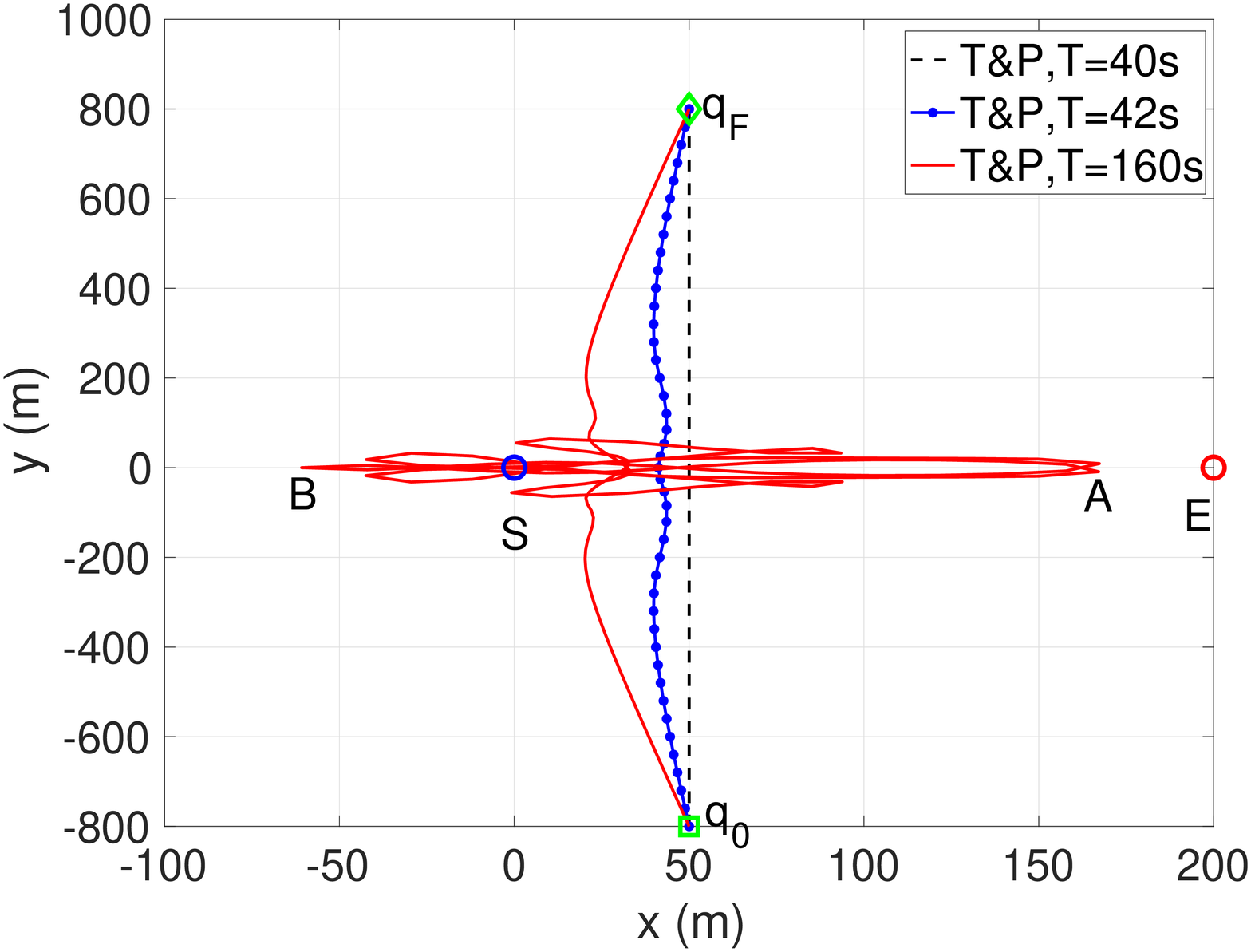}}\subfloat[UAV's Trajectories by NJ/T algorithm.]{\centering{}\includegraphics[width=8cm,height=6cm]{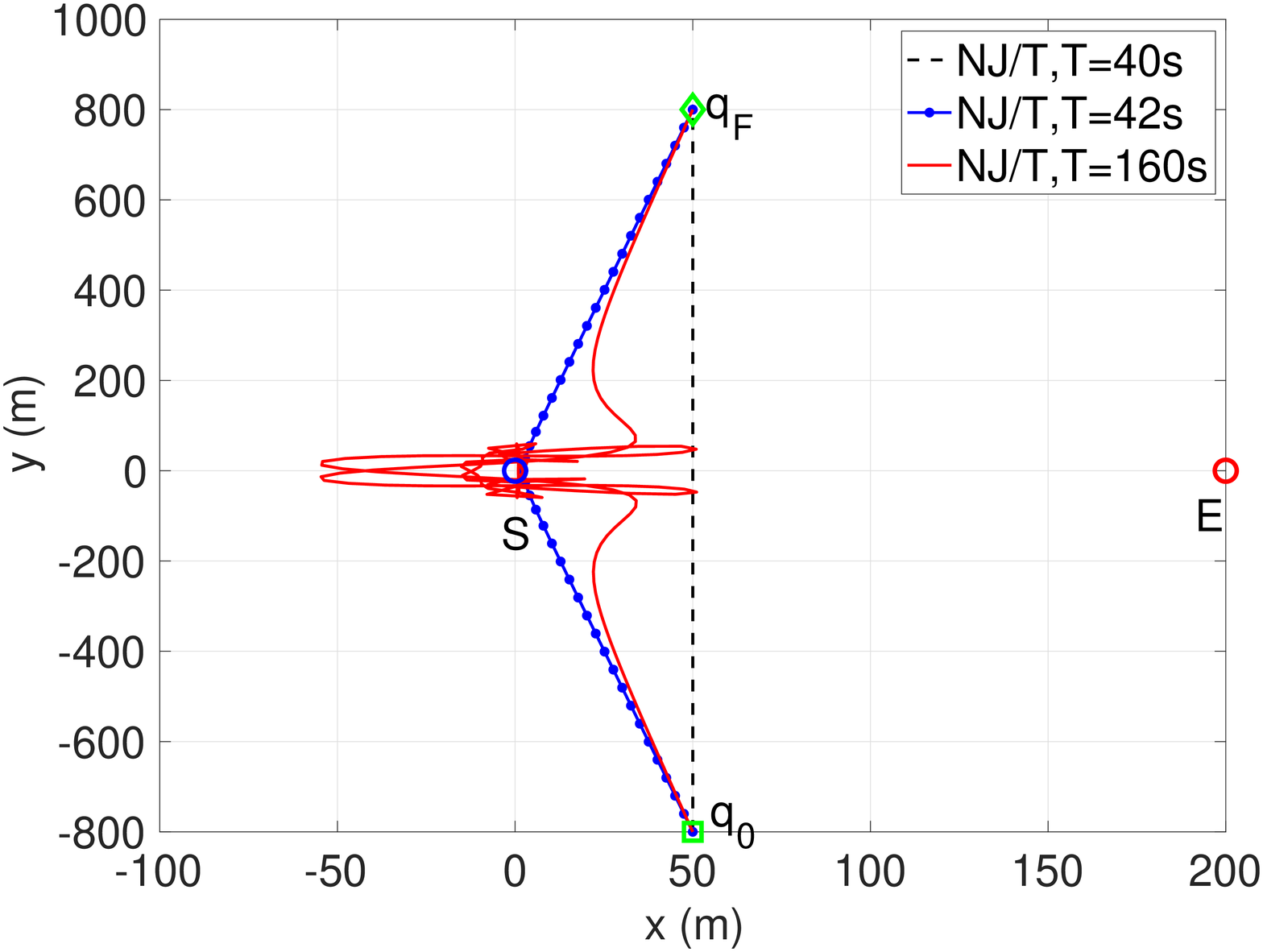}}
\par\end{centering}
\begin{centering}
\subfloat[UAV's Trajectories by NP/T algorithm.]{\centering{}\includegraphics[width=8cm,height=6cm]{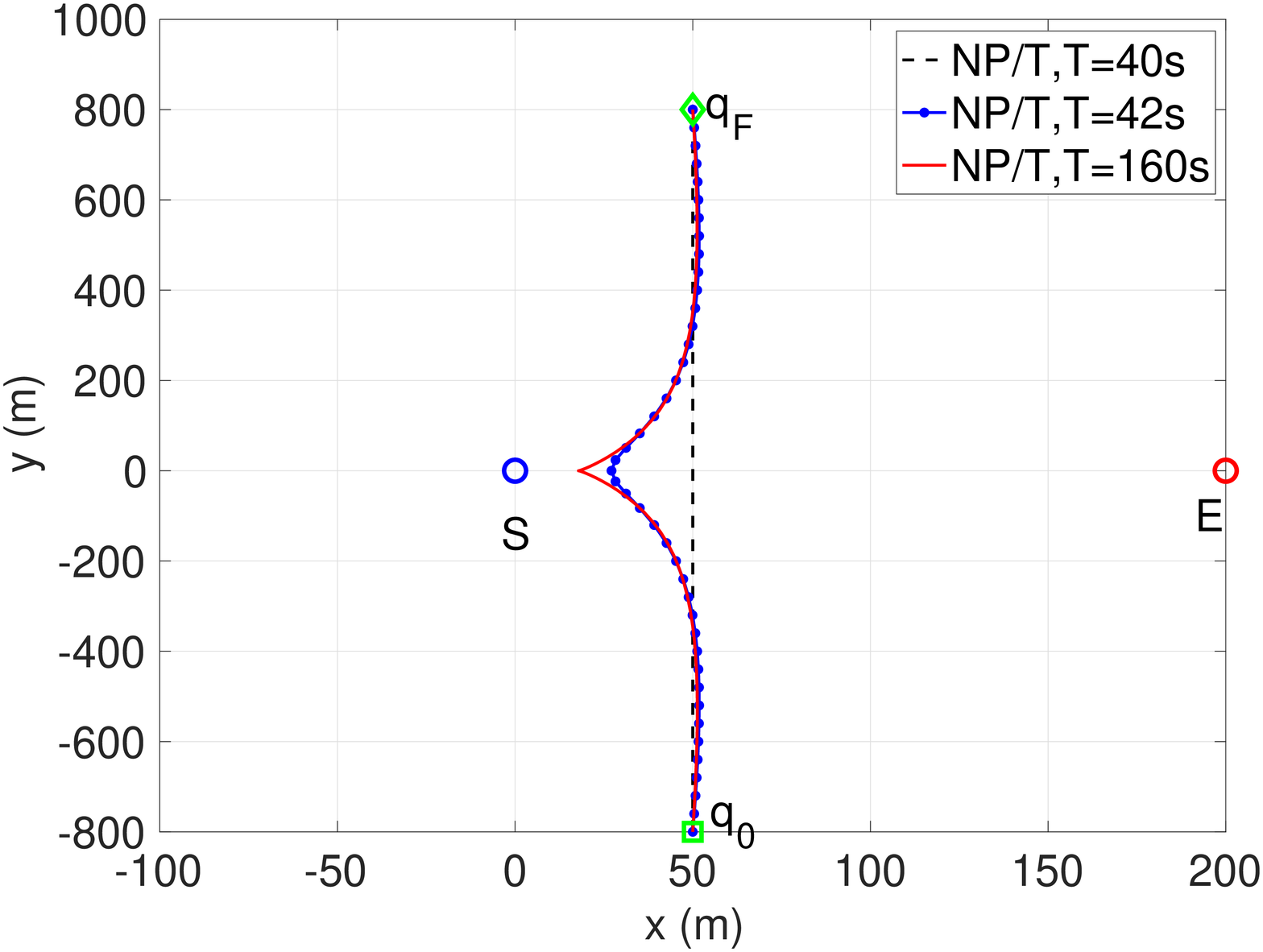}}
\par\end{centering}
\caption{Optimized trajectories of the UAV by different algorithms.}
\end{figure*}

Fig. 2 shows the optimized UAV trajectories by different algorithms
with different values of $T$ when the LIL is $\sigma_{\textrm{RSI}}^{2}=-80$
dBm. For the case with $T=40$ s, the duration is only sufficient
for the UAV to fly from the initial location $\mathbf{q}_{0}$ to
final location $\mathbf{q}_{\textrm{F}}$\emph{ }at speed $V_{\max}$,
thus the trajectories of the three considered algorithms in Fig. 2(a)-(c)
are identical. However, with increased $T$, especially when $T$
is sufficiently large (e.g. $T=160$ s), it is observed that the trajectories
of the three algorithms become significantly different. In particular,
for the proposed P\&T algorithm in Fig. 2(a), the UAV first flies
towards S, then circles around between locations A and B, and finally
reaches $\mathbf{q}_{\textrm{F}}$ by the end of $T$. Along this
optimized trajectory (including both the path and UAV speed), the
UAV can achieve the higher EE for the secrecy communication than that
of other algorithms, since it more efficiently balances between information
reception from S versus jamming signal transmission to E via power\emph{
}control, with less propulsion energy consumption. Specifically, since
the UAV at the location A is far away from S but close to E, the UAV
and S jointly transmit with their maximum powers to ensure a higher
secrecy rate. By contrast, although the UAV at the location B and
S jointly decrease their transmit powers due to the farther distance
from the location B to E, a higher secrecy rate can also be guaranteed
by appropriate power allocation. In Fig. 2(b), since jamming is not
available in NJ/T, the UAV mainly hovers around S to balance the secrecy
rate and the propulsion energy consumption, with only necessary time
left for traveling. For the NP/T algorithm in Fig. 2(c), we observe
that the UAV first reaches a location close to S and then remains
stationary there as long as possible. Despite the high propulsion
energy consumption for remaining stationary, the UAV has to reconcile
a trade-off to obtain the higher secrecy rate, due to the fixed source/UAV
transmit/jamming powers.

\begin{figure*}[t]
\begin{centering}
\subfloat[Energy consumption of the UAV versus $t$.]{\centering{}\includegraphics[width=8cm,height=6cm]{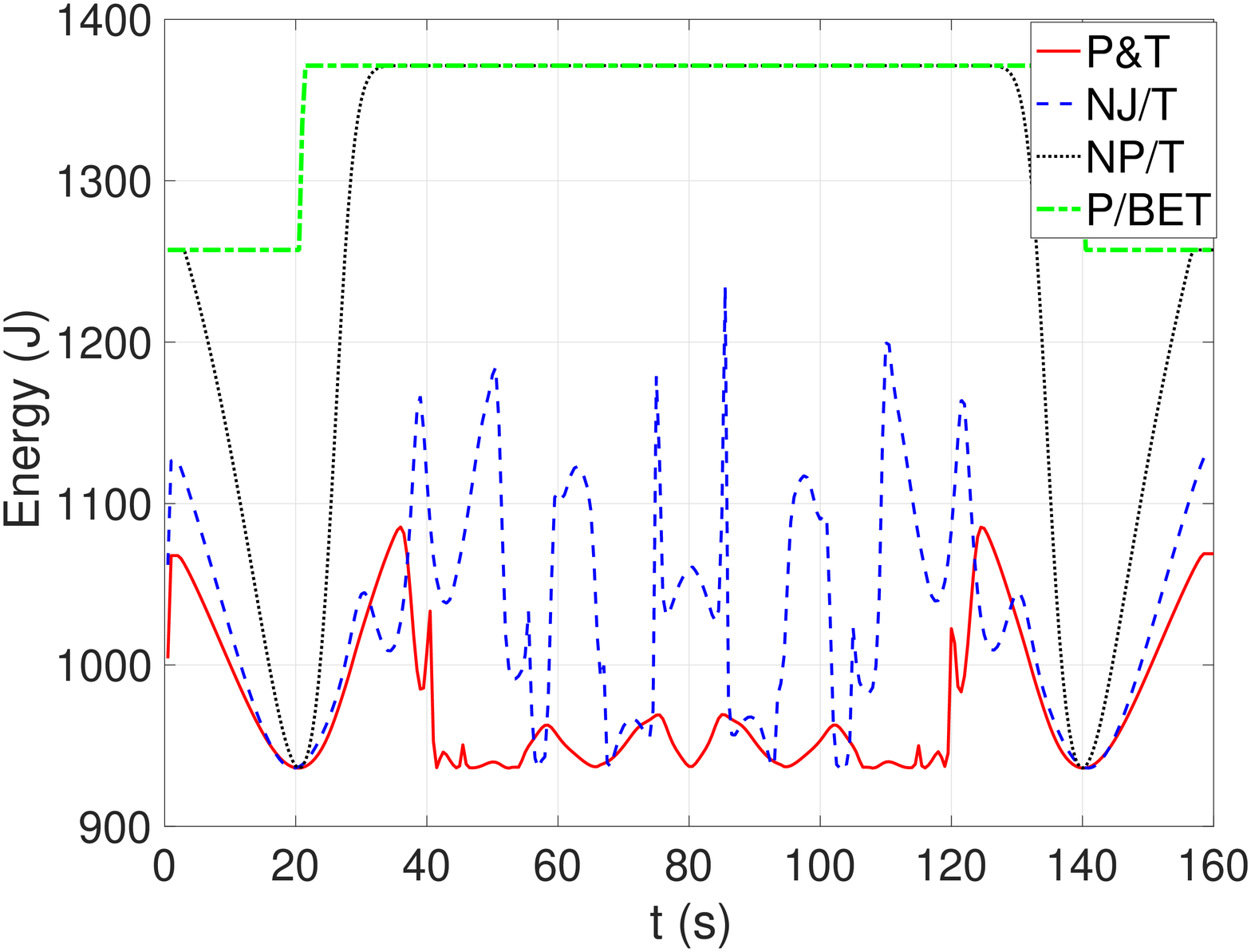}}\subfloat[Speed of the UAV versus $t$.]{\centering{}\includegraphics[width=8cm,height=6cm]{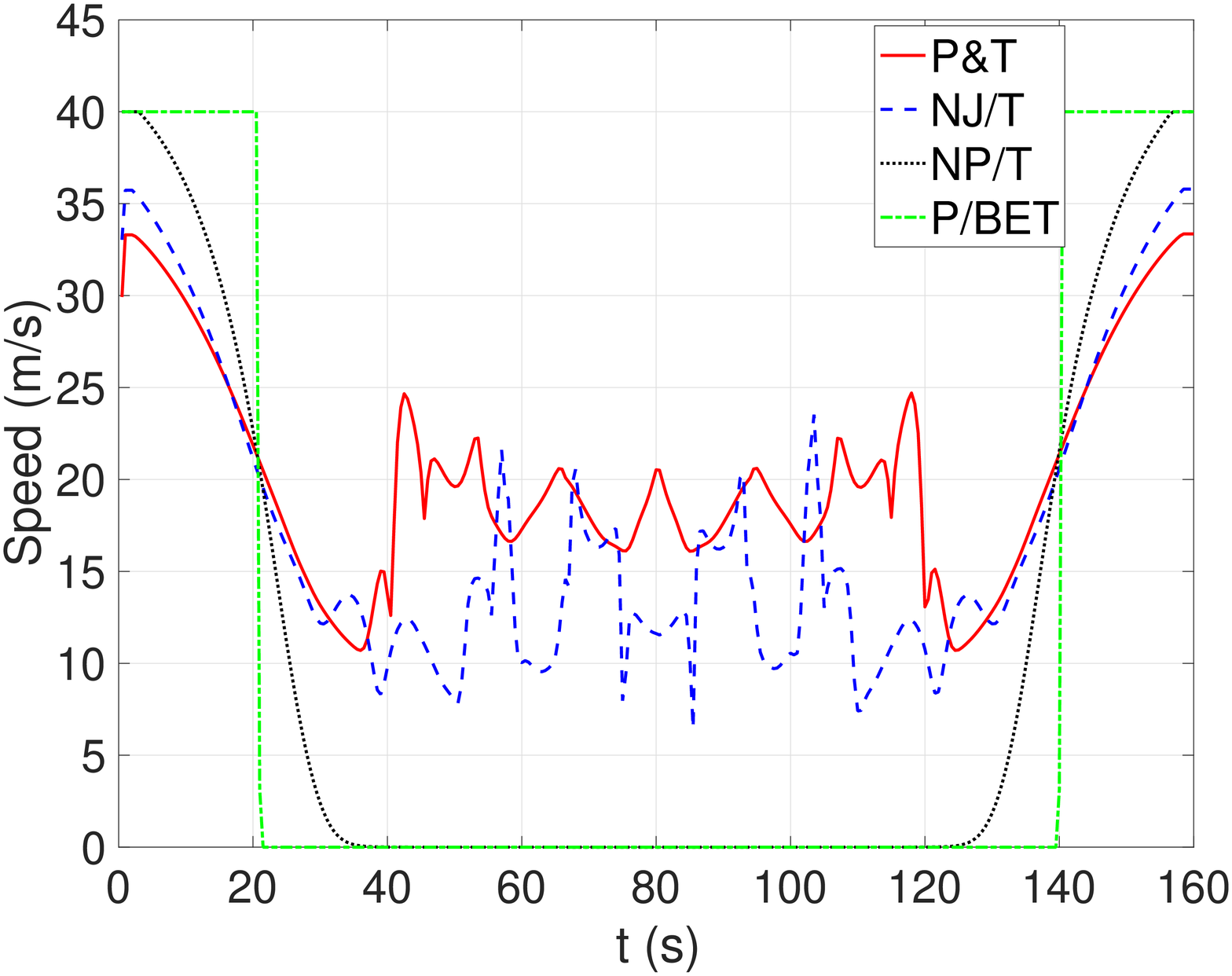}}
\par\end{centering}
\begin{centering}
\subfloat[EE of the UAV versus $T$.]{\centering{}\includegraphics[width=8cm,height=6cm]{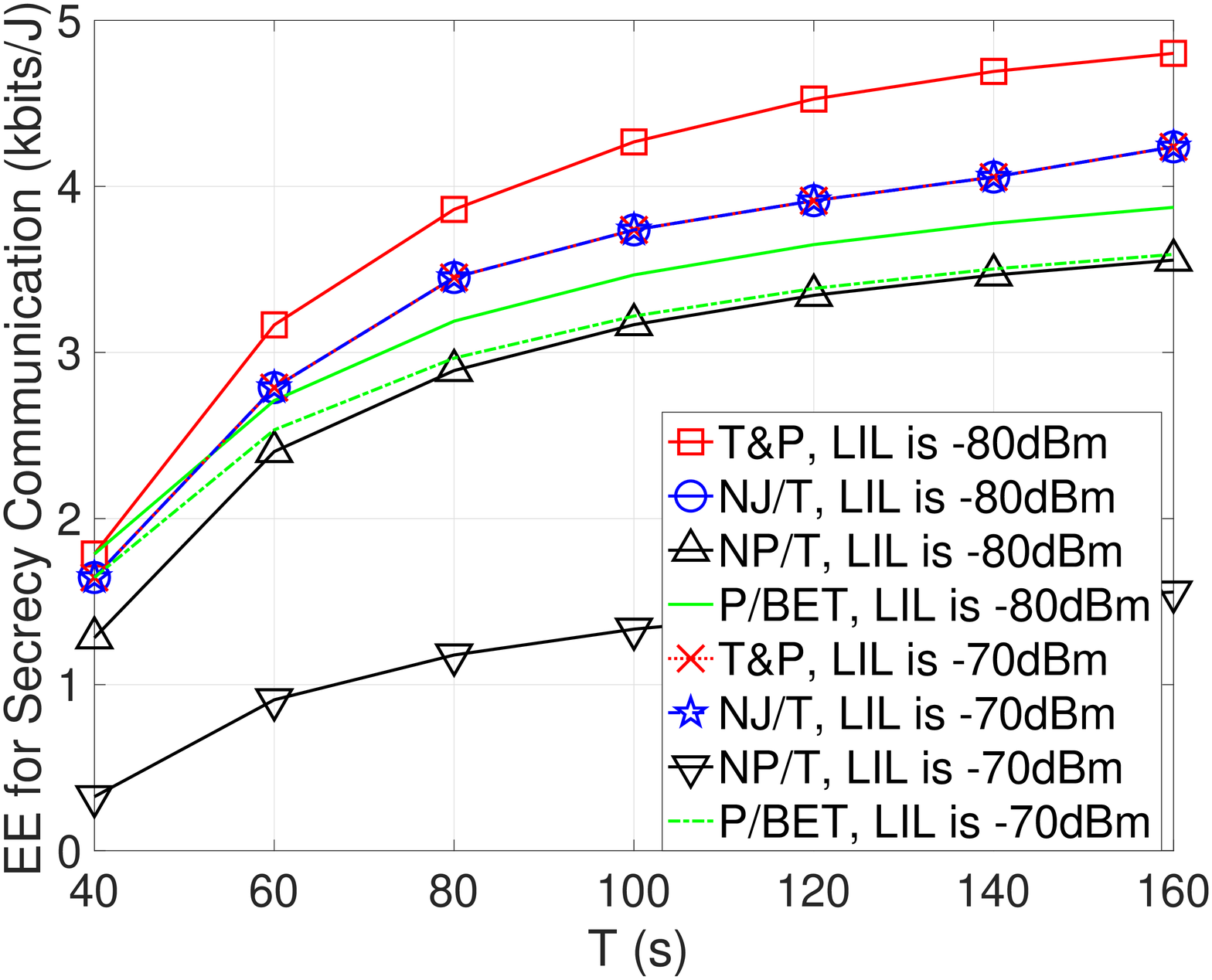}}
\par\end{centering}
\caption{UAV energy consumption, speed and EE for secrecy communication by
different algorithms.}
\end{figure*}

Figs. 3(a) and 3(b) show the UAV energy consumption and speed over
time in $T$, respectively. First, it is observed from Fig. 3(a) that
the UAV in P\&T consumes the least propulsion energy, since along
its optimized trajectory, the UAV can adjust its speed more energy-efficiently
to prevent flying at excessive large and low speeds (see Fig. 3(b)).
By contrast, the energy consumption of the UAV in P/BET algorithm
is highest among all of the other algorithms, due to its heuristic
best-effort trajectory with the maximum speed for flying and zero
speed for hovering. Second, although the UAV in NP/T has a different
hovering location from that in P/BET, they have the same highest energy
consumption when remaining stationary (e.g., from $t=40$ s to 120
s). This indicates that hovering for rotary-wing UAVs is not energy
conserving, which is consistent with \cite{Zeng:2018ux}. Finally,
by comparing Figs. 3(a) and 3(b), we can see that the most energy-efficient
UAV speed is about 20 m/s and the energy consumption increases drastically
when the UAV speed approaches zero.

Fig. 3(c) shows the EE for the secrecy communication versus $T$ under
different values of LIL. It is observed that the EE achieved by all
algorithms increases with $T$ while decreasing with the increase
of LIL, due to the degraded loop channel at the UAV. In particular,
when the value of LIL is -80 dBm, the proposed P\&T algorithm always
outperforms other benchmark algorithms due to its joint optimization
of the trajectory and powers. However, as the LIL value becomes sufficiently
large (e.g., $\sigma_{\textrm{RSI}}^{2}=-70$ dBm), the EE of the
P\&T algorithm reduces to that of the NJ/T algorithm, since it is
ineffective to send jamming signals in this case. Therefore, sending
jamming signals or not from the UAV in P\&T depends mainly on the
level of RSI, and the NJ/T algorithm provides a performance lower
bound for the proposed P\&T algorithm. The above results validate
the potential gain in EE brought by the proposed FD scheme and joint
optimization of transmit/jamming powers and UAV trajectory.

\section{Conclusions}

Security, energy consumption, and spectral efficiency are key factors
for next generation wireless networks with UAVs. Thus, a new FD scheme
for the UAV secrecy communication with EE optimization was proposed
and investigated in this letter. In particular, the EE of a rotary-wing
UAV serving as both a legitimate receiver and a mobile jammer, was
maximized by jointly designing the source/UAV transmit/jamming powers
and the UAV trajectory. An efficient iterative algorithm was proposed
by applying BCD and SCA techniques to solve the problem of the EE
maximization over a given flight period. As compared with the benchmark
schemes without FD transmission, power control or trajectory optimization,
the proposed joint optimization algorithm with FD operation achieves
the highest flexibility in adjusting the UAV jamming power by considering
its practical RSI. Numerical results showed that the EE of UAV secrecy
communication is significantly improved by our proposed algorithm
over the benchmark schemes.

\appendices{}

\bibliographystyle{IEEEtran}
\bibliography{UAV_FD}

\end{document}